\documentclass[12pt]{article}
\pdfoutput=1

\usepackage{a4wide}
\usepackage{amsmath,amssymb}
\usepackage{slashed}
\usepackage{xcolor}
\usepackage{graphicx}
\usepackage{url}
\usepackage{cite}
\usepackage[colorlinks=true,allcolors=darkred,pdfborder={0 0 0},linktocpage=false]{hyperref}

\definecolor{darkred}{rgb}{0.6,0,0}
\def\hc{\text{h.c.}}

\begin{document}

\vspace*{-2cm}
\begin{flushright}
IFIC/18-11 \\
\vspace*{2mm}
\today
\end{flushright}

\begin{center}
\vspace*{15mm}

\vspace{1cm}
{\Large \bf 
Anomalies in $\boldsymbol{b \to s}$ transitions and dark matter
} \\
\vspace{1cm}

{\bf Avelino Vicente}

 \vspace*{.5cm} 
Instituto de F\'{\i}sica Corpuscular (CSIC-Universitat de Val\`{e}ncia), \\
Apdo. 22085, E-46071 Valencia, Spain

\end{center}

\vspace*{10mm}
\begin{abstract}\noindent\normalsize
Since 2013, the LHCb collaboration has reported on the measurement of
several observables associated to $b \to s$ transitions, finding
various deviations from their predicted values in the Standard
Model. These include a set of deviations in branching ratios and
angular observables, as well as in the observables $R_K$ and
$R_{K^\ast}$, specially built to test the possible violation of Lepton
Flavor Universality. Even though these tantalizing hints are not
conclusive yet, the $b \to s$ anomalies have gained considerable
attention in the flavor community. Here we review New Physics models
that address these anomalies and explore their possible connection to
the dark matter of the Universe. After discussing some of the ideas
introduced in these works and classifying the proposed models, two
selected examples are presented in detail in order to illustrate the
potential interplay between these two areas of current particle
physics.
\end{abstract}

\newpage

\tableofcontents

\newpage

\section{Introduction} \label{sec:intro}

The Standard Model (SM) of particle physics provides an excellent
description for a vast amount of phenomena and can be regarded as one
of the most successful scientific theories ever built. In fact, with
the recent discovery of the last missing piece, the Higgs boson, the
particle spectrum is finally complete and the SM looks stronger than
ever. However, and despite its enormous success, there are several
indications that clearly point towards the existence of a more
complete theory, with neutrino masses and the baryon asymmetry of the
Universe as the most prominent examples.

Another open question is the nature of the dark matter (DM) that
accounts for $27 \%$ of the energy density of the Universe
\cite{Ade:2015xua}. Several ideas have been proposed to address this
fundamental problem in current physics. Under the hypothesis that the
DM is composed of particles, these cannot be identified with any of
the states in the SM, hence demanding an extension of the model with
new states and, possibly, new dynamics. Again, many directions
exist. Interestingly, in scenarios involving New Physics (NP) at the
TeV scale, the first signals from the new DM sector might be found in
experiments not specially designed to look for them.

Rare decays stand among the most powerful tests of the SM. Since 2013,
results obtained by the LHCb collaboration have led to an increasing
interest in B physics, particularly in processes involving $b \to s$
transitions. Deviations from the SM expectations have been reported in
several observables, some of them hinting at the violation of lepton
flavor universality (LFU), a central feature in the SM. Even though
these anomalies could be caused by a combination of unfortunate
fluctuations and, perhaps, a poor theoretical understanding of some
processes, it is tempting to speculate about their possible origin in
terms of NP models, in particular models linking them to other open
problems.

This {\it mini-review} will pursue this goal, focusing on NP scenarios
that relate the $b \to s$ anomalies to DM. Several works
\cite{Sierra:2015fma,Belanger:2015nma,Allanach:2015gkd,Bauer:2015boy,Celis:2016ayl,Altmannshofer:2016jzy,Ko:2017quv,Ko:2017yrd,Cline:2017lvv,Sala:2017ihs,Ellis:2017nrp,Kawamura:2017ecz,Baek:2017sew,Cline:2017aed,Cline:2017qqu,Dhargyal:2017vcu,Chiang:2017zkh,Falkowski:2018dsl,Arcadi:2018tly}
have already explored this connection, mostly by means of specific
models that accommodate the observations in $b \to s$ transitions with
a new dark sector. We will review some of the ideas introduced in
these works and highlight those that deserve further exploration. We
will also classify the proposed models into two general categories:
(i) models in which the NP contributions to $b \to s$ transitions and
DM production in the early Universe share a common mediator, and (ii)
models with the DM particle running in loop diagrams that contribute
to the solution of the $b \to s$ anomalies. After a general
discussion, a selected example of each class will be presented in
detail.

The rest of the manuscript is organized as follows. First, we review
the anomalies in $b \to s$ transitions in Sec. \ref{sec:anomalies} and
interpret the experimental results in a model independent way in
Sec. \ref{sec:interpretation}. In Sec. \ref{sec:link} we discuss and
classify the proposed New Physics explanations to these anomalies that
involve a link to the dark matter problem. Secs. \ref{sec:model1} and
\ref{sec:model2} present two simple example models that illustrate
this connection. Finally, we summarize and draw our conclusions in
Sec. \ref{sec:conclusions}.

\section{Experimental situation} \label{sec:anomalies}

We begin by discussing the present experimental situation. The observed
anomalies in $b \to s$ transitions can be classified into two classes:
(1) branching ratios and angular observables, and (2) lepton flavor
universality violating (LFUV) anomalies. Although they might be
related (and caused by the same NP), they are conceptually different.

\vspace*{0.2cm}

{\bf Branching ratios and angular observables:} using state-of-the-art
computations of the hadronic form factors involved, one can compute
branching ratios and angular observables for $b \to s$ processes such
as $B \to K^{\ast} \ell^+ \ell^-$ and look for deviations from the SM
predictions. For the comparison to be meaningful, one must have a good
knowledge of all possible Quantum ChromoDynamics (QCD) effects that
might pollute the theoretical calculation and we currently have at our
disposal several methods to minimize or at least estimate the
uncertainties.~\footnote{The size of the hadronic uncertainties in
  different calculations is a matter of hot debate nowadays. We will
  not discuss this issue here but just refer to the recent studies
  regarding form factors \cite{Jager:2012uw,Descotes-Genon:2014uoa}
  and non-local contributions
  \cite{Khodjamirian:2010vf,Ciuchini:2015qxb,Capdevila:2017ert,Chobanova:2017ghn,Bobeth:2017vxj,Blake:2017fyh}
  for extended discussions.} In particular, a basis of optimized
observables for the decay $B \to K^\ast \mu^+ \mu^-$, specially
designed to reduce the hadronic uncertainties, was introduced in
\cite{Descotes-Genon:2013vna}. In 2013, the LHCb collaboration
published results on these observables using their $1$ fb$^{-1}$
dataset, finding a $3.7 \sigma$ deviation between the measurement and
the SM prediction for the $P_5^\prime$ angular observable in one
dimuon invariant mass bin \cite{Aaij:2013qta}. A systematic deficit
with respect to the SM predictions in the branching ratios of several
processes, mainly $B_s \to \phi \mu^+ \mu^-$, was also reported by
LHCb \cite{Aaij:2013aln}. These discrepancies have been found later in
other datasets. In 2015, LHCb confirmed these anomalies using their
full Run 1 dataset with $3$
fb$^{-1}$~\cite{Aaij:2015oid,Aaij:2015esa}, whereas in 2016 the Belle
collaboration presented an independent measurement of $P_5^\prime$,
compatible with the LHCb
result~\cite{Abdesselam:2016llu,Wehle:2016yoi}. More recently, both
ATLAS \cite{ATLAS:2017dlm} and CMS \cite{CMS:2017ivg} have also
presented preliminary results on the $B \to K^\ast \mu^+ \mu^-$
angular observables, with relatively good agreement with LHCb.

\vspace*{0.2cm}

{\bf LFUV anomalies:} one of the central features of the SM is that
gauge bosons couple with the same strength to all three families of
leptons. This prediction can be tested by measuring observables such
as the $R_{K^{(\ast)}}$ ratios, defined as~\cite{Hiller:2003js}
\begin{align}
R_{K^{(\ast)}} = \frac{ \Gamma(B \rightarrow K^{(\ast)} \mu^+ \mu^-)}{\Gamma(B \rightarrow K^{(\ast)} e^+ e^-)} \, ,
\end{align}
measured in specific dilepton invariant mass squared ranges $q^2 \in
[q^2_{\rm min}, q^2_{\rm max}]$. In the SM, these ratios should be
very approximately equal to one. Furthermore, hadronic uncertainties
are expected to cancel to very good approximation in these ratios,
which implies that, in contrast to the previous class of anomalies,
deviations in these observables cannot be explained by uncontrolled
QCD effects and would be a clear indication of NP at work. For this
reason, they are sometimes referred to as {\it clean
  observables}. Interestingly, in 2014 the LHCb collaboration measured
$R_K$ in the region $[1,6]$~GeV$^2$~\cite{Aaij:2014ora}, finding a
value significantly lower than one, while in 2017 similar measurements
of the $R_{K^\ast}$ ratio in two $q^2$ bins \cite{Aaij:2017vbb} were also
found to depart from their SM expected values:
\begin{align}
R_K &= 0.745^{+0.090}_{-0.074}\pm0.036    \,, \quad
q^2 \in [1,6]~\text{GeV}^2 \,, \nonumber \\[0.2cm]
R_{K^\ast} &= 0.660^{+0.110}_{-0.070}\pm0.024    \,, \quad
q^2 \in [0.045,1.1]~\text{GeV}^2\,, \nonumber \\[0.2cm]
R_{K^\ast} &= 0.685^{+0.113}_{-0.069}\pm0.047    \,, \quad
q^2 \in [1.1,6.0]~\text{GeV}^2 \,. 
\end{align}
The comparison between these experimental results and the SM predictions~\cite{Descotes-Genon:2015uva,Bordone:2016gaq},
\begin{align}
R_K^{\rm{SM}} &= 1.00 \pm 0.01   \,, \quad
q^2 \in [1,6]~\text{GeV}^2 \,, \nonumber \\[0.2cm]
R_{K^\ast}^{\rm{SM}} &= 0.92 \pm 0.02    \,, \quad
q^2 \in [0.045,1.1]~\text{GeV}^2 \,,\nonumber \\[0.2cm]
R_{K^\ast}^{\rm{SM}} &= 1.00 \pm 0.01    \,, \quad
q^2 \in [1.1,6.0]~\text{GeV}^2 \,,
\end{align}
shows deviations from the SM at the $2.6\,\sigma$ level in the case of
$R_K$, $2.2\,\sigma$ for $R_{K^\ast}$ in the low-$q^2$ region, and
$2.4\,\sigma$ for $R_{K^\ast}$ in the central-$q^2$ region. Finally,
Belle has recently measured the LFUV observable $Q_5 = P_5^{\mu \,
  \prime} - P_5^{e \, \prime}$, with the observable $P_5^{e \, \prime}$
defined for $B \to K^\ast e^+ e^-$ analogously to $P_5^{\mu \, \prime}
\equiv P_5^\prime$ for $B \to K^\ast \mu^+
\mu^-$~\cite{Capdevila:2016ivx}. The result, although statistically
not very significant, also points towards the violation of LFU
\cite{Wehle:2016yoi}.

\vspace*{0.2cm}

Summarizing, there are at present two sets of experimental anomalies
in processes involving $b \to s$ transitions at the quark level. While
the relevance of the first set is currently a matter of discussion due
to the possibility of unknown QCD effects faking the deviations from
the SM, the second can only be explained by NP violating LFU. In
principle, these two classes of anomalies can be completely unrelated
but, as we will see in the next Section, global analyses of all
experimental data in $b \to s$ transitions indicate that a common
explanation (in terms of a single effective operator) can address both
sets in a satisfactory and economical way. This intriguing result has
made the $b \to s$ anomalies a topic of great interest currently.

Finally, it is very interesting to note the existence of an
independent set of anomalies in $b \to c$ transitions. Several
experimental measurements of the ratios $R(D)$ and $R(D^\ast)$ have
been found to depart from their SM predictions, with a global
discrepancy at the $\sim 4 \sigma$ level~\cite{RDRDstar}. Recently,
the $R(J/\psi)$ ratio has also been measured by the LHCb
collaboration, finding again a deviation from the SM expected value
\cite{Aaij:2017tyk}. Compared to the $b \to s$ anomalies, the $b \to
c$ anomalies are of a different nature and, if real, they could have a
completely different origin. For instance, they would involve a new
charged current, instead of a neutral one, hence requiring the new
mediators to be much lighter to be able to compete with the SM $W$
boson tree-level exchange. However, many authors have proposed models
that can simultaneously address both sets of anomalies. We refer to
\cite{Buttazzo:2017ixm} for a general discussion on combined
explanations and ignore the $b \to c$ anomalies for the rest of this
paper.

\section{Model independent interpretation} \label{sec:interpretation}

The experimental tensions discussed in the previous Section must be
properly quantified and interpreted. {\bf Quantification} is crucial
to determine whether the anomalies can be explained by fluctuations in
the data or they truly indicate a statistically significant deviation
from the SM. Assuming that these tensions are caused by genuine NP,
the ultimate goal is to construct a specific model in which they are
solved. However, the first step in this direction must be a {\bf model
  independent interpretation} of the experimental data in order to
identify the ingredients that this new scenario must include. This is
achieved by adopting an approach based on effective operators, valid
under the assumption that all NP degrees of freedom lie at energies
well above the relevant energy scales for the observables of interest.

The effective Hamiltonian for $b \to s$ transitions is usually
written as
\begin{equation} \label{eq:effH}
\mathcal H_{\text{eff}} = - \frac{4 G_F}{\sqrt{2}} \, V_{tb} V_{ts}^\ast \, \frac{e^2}{16 \pi^2} \, \sum_i \left(C_i \, \mathcal O_i + C^\prime_i \, \mathcal O^\prime_i \right) + \hc \, .
\end{equation}
Here $G_F$ is the Fermi constant, $e$ the electric charge and $V$ the
Cabibbo-Kobayashi-Maskawa (CKM) matrix. $\mathcal O_i$ and $\mathcal
O^\prime_i$ are the effective operators that contribute to $b \to s$
transitions, and $C_i$ and $C^\prime_i$ their Wilson coefficients. The
most relevant operators for the interpretation of the $b \to s$
anomalies are
\begin{align}
\mathcal O_9 &= \left( \bar s \gamma_\mu P_L b \right) \, \left( \bar \ell \gamma^\mu \ell \right) \, ,   &   \mathcal O^\prime_9 &= \left( \bar s \gamma_\mu P_R b \right) \, \left( \bar \ell \gamma^\mu \ell \right) \, , \label{eq:O9} \\
\mathcal O_{10} &= \left( \bar s \gamma_\mu P_L b \right) \, \left( \bar \ell \gamma^\mu \gamma_5 \ell \right) \, ,   &   \mathcal O^\prime_{10} &= \left( \bar s \gamma_\mu P_R b \right) \, \left( \bar \ell \gamma^\mu \gamma_5 \ell \right) \, . \label{eq:O10}
\end{align}
Here $\ell = e, \mu, \tau$. In fact, the operators and Wilson
coefficients carry flavor indices and we are omitting them to simplify
the notation. When necessary, we will denote a particular lepton flavor
with a superscript, e.g. $C_9^\mu$ and $\mathcal O_9^\mu$, for
muons. It is also convenient to split the Wilson coefficients in two
pieces: the SM contributions and the NP contributions,
defining~\footnote{Similar splittings could be defined for the Wilson
  coefficients of the primed operators, $C^\prime_9$ and
  $C^\prime_{10}$, but in this case the SM contributions are suppresed
  and one has $C^\prime_9 \simeq C^{\prime \: \text{NP}}_9$ and
  $C^\prime_{10} \simeq C^{\prime \: \text{NP}}_{10}$.}
\begin{eqnarray}
C_9 &=& C_9^{\text{SM}} + C_9^{\text{NP}} \, , \\
C_{10} &=& C_{10}^{\text{SM}} + C_{10}^{\text{NP}} \, .
\end{eqnarray}
The SM contributions have been computed at NNLO at $\mu_b = 4.8$ GeV,
obtaining $C_9^{\text{SM}}(\mu_b) = 4.07$ and
$C_{10}^{\text{SM}}(\mu_b) = - 4.31$ (see
\cite{Descotes-Genon:2013vna} and references therein), leaving the NP
contributions as parameters to be determined (or at least constrained)
by using experimental data.

It is in principle possible to derive limits for the NP contributions
considering each observable independently, but this approach would
completely miss the global picture. The effective operators in
Eq. \eqref{eq:effH} contribute to several observables and one expects
the presence of NP to be revealed by a pattern of deviations from the
SM expectations, rather than by a single anomaly. For this reason,
global fits constitute the best approach to analyze the available
experimental data. Interestingly, several independent fits
\cite{Capdevila:2017bsm,Altmannshofer:2017yso,DAmico:2017mtc,Hiller:2017bzc,Geng:2017svp,Ciuchini:2017mik,Alok:2017sui,Hurth:2017hxg}
have found a remarkable tension between the SM and experimental data
on $b \to s$ transitions which is clearly reduced with the addition of
NP contributions. Although the numerical details (such as statistical
significances) differ among different analyses, there is a general
consensus on the following qualitative results:

\begin{itemize}
\item Global fits improve substantially with a negative contribution
  in $C_9^{\mu , \text{NP}}$, with $C_9^{\mu , \text{NP}} \sim - 25 \%
  \times C_9^{\mu , \text{SM}}$, leading to a total Wilson coefficient
  $C_9^\mu$ significantly smaller than the one in the SM.
\item NP contributions in other Wilson coefficients can also improve
  the fit, but only in a sub-dominant way. For instance, the anomalies
  can also be accommodated in scenarios with $C_9^{\mu , \text{NP}} =
  - C_{10}^{\mu , \text{NP}}$ or $C_9^{\mu , \text{NP}} = -
  C_{9}^{\prime \: \mu , \text{NP}}$, without a clear statistical
  preference with respect to the scenario with NP only in
  $C_9^\mu$.~\footnote{Such patterns for the Wilson coefficients are
    automatically obtained if the NP states couple to SM fermions with
    specific chiralities. For instance, the relation $C_9^{\mu ,
      \text{NP}} = - C_{10}^{\mu , \text{NP}}$ is obtained in models
    where the NP states only couple to the \textit{left-handed}
    muons. Two examples of this class of models are shown in
    Secs. \ref{sec:model1} and \ref{sec:model2}.}
\item Other operators involving muons are perfectly compatible with
  their SM values. Similarly, no NP is required for operators
  involving electrons or tau leptons.
\end{itemize}

Armed with these results, model builders can construct specific models
where all requirements are met and the anomalies explained. Similarly,
one can extract interesting implications for model building by
explaining the anomalies in terms of gauge invariant effective
operators, see \cite{Celis:2017doq} for a recent analysis. Either way,
the resulting profile of NP contributions reveals a pattern that was
not predicted by any theoretical framework, such as supersymmetry, and
many new models have been put forward. In the next Section we will
discuss some of these models, in particular those linking the $b \to
s$ anomalies to dark matter.

\section{Linking the $\boldsymbol{b \to s}$ anomalies to dark matter} \label{sec:link}

After discussing the current experimental situation in $b \to s$
transitions, let us focus on possible connections to the dark matter
problem. These have been explored in
\cite{Sierra:2015fma,Belanger:2015nma,Allanach:2015gkd,Bauer:2015boy,Celis:2016ayl,Altmannshofer:2016jzy,Ko:2017quv,Ko:2017yrd,Cline:2017lvv,Sala:2017ihs,Ellis:2017nrp,Kawamura:2017ecz,Baek:2017sew,Cline:2017aed,Cline:2017qqu,Dhargyal:2017vcu,Chiang:2017zkh,Falkowski:2018dsl,Arcadi:2018tly}. In
general, the proposed models that solve the $b \to s$ anomalies and
explain the origin of the dark matter of the Universe can be
classified into two principal categories:

\begin{itemize}
\item {\bf Portal models:} models in which the mediator responsible
  for the NP contributions to $b \to s$ transitions also mediates the
  DM production in the early Universe.
\item {\bf Loop models:} models that induce the required NP
  contributions to $b \to s$ transitions with loops containing the DM
  particle.
\end{itemize}

In the case of {\bf portal models}, the usual scenario considers a
$U(1)$ gauge extension of the SM that leads to the existence of a new
massive gauge boson after spontaneous symmetry breaking. The resulting
$Z^\prime$ boson induces a new neutral current contribution in $b \to
s$ transitions and mediates the production of DM particles in the
early Universe via a $Z^\prime$ portal interaction. This setup was
first considered in \cite{Sierra:2015fma}. In this particular
realization of the general idea, the SM fermions were assumed to be
neutral under the new $U(1)_X$ gauge symmetry and the $Z^\prime$
couplings to quarks ($\overline b s$) and leptons ($\mu^+ \mu^-$),
necessary to explain the $b \to s$ anomalies, are generated at
tree-level via mixing with new vector-like (VL)
fermions. Additionally, the $Z^\prime$ boson also couples to the
scalar field $\chi$, the DM candidate in this model, automatically
stabilized by an remnant $\mathbb{Z}_2$ symmetry after $U(1)_X$
breaking. This model will be reviewed in more detail in
Sec. \ref{sec:model1}. Variants of this setup with fermionic DM also
exist. In \cite{Celis:2016ayl}, a horizontal $U(1)_{B_1+B_2-2 B_3}$
gauge symmetry is introduced, with $B_i$ the baryon number of the ith
fermion family. The resulting $Z^\prime$ boson couples directly to the
SM quarks, while the coupling to muons is obtained by introducing a VL
lepton. This allows to accommodate the anomalies in $b \to s$
transitions. Furthermore, the model also contains a Dirac fermion that
is stable due to a remnant $\mathbb{Z}_2$ symmetry, in a similar
fashion as in \cite{Sierra:2015fma}, which becomes the DM
candidate. Similarly, Ref. \cite{Altmannshofer:2016jzy} builds on the
well-known $U(1)_{L_\mu-L_\tau}$ model of \cite{Altmannshofer:2014cfa}
and extends it to include a stable Dirac fermion with a relic density
also determined by $Z^\prime$ portal interactions, while
\cite{Arcadi:2018tly} considers a similar model but makes use of
kinetic mixing between the $Z^\prime$ and the SM neutral gauge
bosons. Ref. \cite{Falkowski:2018dsl} considers vector-like neutrino
DM in a setup analogous to \cite{Sierra:2015fma} extended with
additional VL fermions. Ref. \cite{Cline:2017lvv} explored a pair of
scenarios based on a $U(1)$ gauge symmetry supplemented with VL
fermions and a fermionic DM candidate, of Dirac or Majorana
nature. This paper focuses on effects in indirect detection
experiments, aiming at an explanation of the excess of events in
antiproton spectra reported by the AMS-02 experiment in 2016
\cite{Aguilar:2016kjl}. Other works that adopt the standard $Z^\prime$
portal setup are \cite{Allanach:2015gkd,Ellis:2017nrp}. Finally,
\cite{Sala:2017ihs} considers a light mediator (not the usual heavy
$Z^\prime$) that contributes to $b \to s$ transitions and decays
predominantly into invisible final states, possibly made of light DM
particles.

It is important to note that the phenomenology of these $Z^\prime$
portal models differs substantially from the standard $Z^\prime$
portal phenomenology. This is due to the fact that the $Z^\prime$
bosons in these models couple with different strengths to different
fermion families, as required to accommodate the LFUV hints observed
by the LHCb collaboration ($R_K$ and $R_{K^\ast}$). For instance, DM
annihilation typically yields muon and tau lepton pairs, but not
electrons and positrons. Direct detection experiments are also more
challenging that in the standard $Z^\prime$ portal scenario, since the
DM candidate typically does not couple to first generation quarks,
more abundant in the nucleons.

\vspace*{0.5cm}

In what concerns {\bf loop models}, many variations are possible. To
the best of our knowledge, the first model of this type that appeared
in the literature is \cite{Kawamura:2017ecz}, based on previous work
on loop models for the $b \to s$ anomalies, without connecting to the
DM problem, in \cite{Gripaios:2015gra,Arnan:2016cpy}. In this model
the SM particle content is extended with two VL pairs of $SU(2)_L$
doublets, with the same quantum numbers as the SM quark and lepton
doublets, but charged under a global $U(1)_X$ symmetry. The model also
contains the complex scalar $X$, singlet under the SM gauge symmetry
and also charged under $U(1)_X$. With these states, one can draw a
1-loop diagram contributing to the $b \to s$ observables relevant to
explain the anomalies. Furthermore, if the global $U(1)_X$ is
conserved, the lightest $U(1)_X$-charged state becomes stable. In this
work, this state is assumed to be $X$, hence the DM candidate in the
model. A more detail discussion about this model can be found in
Sec. \ref{sec:model2}. Two similar setups can be found in
\cite{Chiang:2017zkh}, where a different set of global symmetries are
considered ($U(1) \times \mathbb{Z}_2$ and $U(1) \times \mathbb{Z}_3$)
in order to stabilize a scalar DM candidate. This paper also includes
right-handed neutrinos in order to accommodate non-zero neutrino
masses with the type-I seesaw mechanism and explores the lepton flavor
violating phenomenology of the model in detail. A Majorana fermionic
DM candidate was considered in \cite{Cline:2017qqu}.  Similarly to the
previously mentioned models, this scenario also addresses the $b \to
s$ anomalies at 1-loop level introducing a minimal number of fields:
just a VL quark ($\Psi$) and an inert scalar doublet ($\Phi$), in
addition to the fermion singlet that constitutes the DM candidate. The
model is supplemented with a discrete $\mathbb{Z}_2$ symmetry to
ensure the stability of the DM particle. Interestingly, the model can
be tested in direct DM detection experiments as well as at the Large
Hadron Collider (LHC), where the states $\Psi$ and $\Phi$ can be
pair-produced and lead to final states with hard leptons and missing
energy. Finally, an extended loop model for the $b \to s$ anomalies
which also has an additional $U(1)$ gauge symmetry, contains a scalar
DM candidate and explains neutrino masses can be found in
\cite{Dhargyal:2017vcu}.

\vspace*{0.5cm}

Finally, let us comment on other models and works that do not easily
fit within any of the two categories mentioned above. The model in
\cite{Belanger:2015nma} is very similar to the model in
\cite{Sierra:2015fma}. It also extends the SM with a complex scalar,
VL quarks and leptons and a new $U(1)_X$ gauge symmetry that breaks
down to a $\mathbb{Z}_2$ parity. However, the VL leptons carry
different $U(1)_X$ charges, leading to a loop-induced $Z^\prime \mu^+
\mu^-$ coupling. This changes the DM phenomenology dramatically. The
dominant mechanism for the DM production in the early Universe is not
a $Z^\prime$ portal interaction, but t-channel exchange of VL
leptons. The model in \cite{Ko:2017yrd} can be regarded as a {\it
  hybrid} model, with features from both portal and loop models. The
SM symmetry group is extended with a new $U(1)_{\mu - \tau} \times
\mathbb{Z}_2$ piece. The first factor leads to the existence of a
massive $Z^\prime$ boson while the second one stabilizes a scalar DM
candidate. The $Z^\prime \overline{b} s$ coupling is generated with a
loop containing the $\mathbb{Z}_2$-odd fields and the dominant DM
production mechanism is a $Z^\prime$ portal interaction, mainly with
leptons. The $Z^\prime \overline{b} s$ coupling is also loop-generated
in \cite{Baek:2017sew}, but in this case production of DM particles
takes place via a Higgs portal. \cite{Ko:2017quv} proposes an extended
Scotogenic model for neutrino masses \cite{Ma:2006km} supplemented
with a non-universal $U(1)$ gauge group. The DM candidate in this case
is the lightest fermion singlet and is produced by Yukawa
interactions. Finally, two models that address the $b \to s$ anomalies
with leptoquarks and include DM candidates were introduced in
\cite{Bauer:2015boy} and \cite{Cline:2017aed}. In the former the DM
candidate is a component of an $SU(2)_L$ multiplet introduced to
enhance the diphoton rate of a scalar in the model, whereas in the
latter the DM candidate is a baryon-like composite state in a model
with strong dynamics at the TeV scale.

\vspace*{0.5cm}

Having reviewed and classified the proposed models, we now proceed to
discuss in some detail two specific examples. These illustrate the
main features of portal and loop models.

\section{An example portal model} \label{sec:model1}

We will now review the model introduced in \cite{Sierra:2015fma},
arguably one of the simplest scenarios to account for the $b \to s$
anomalies with a dark sector. Some of the ingredients of this model
were already present in the model of \cite{Altmannshofer:2014cfa},
which is extended in the quark sector (following the same lines as in
the lepton sector). It also includes a dark matter candidate that
couples to the SM fields via the same mediator that leads to an
explanation to the $b \to s$ anomalies, a heavy $Z^\prime$ boson. A
variation of this scenario with a loop-induced coupling to muons
appeared afterwards in \cite{Belanger:2015nma}, whereas the
phenomenology of an extension to account for neutrino masses is
discussed in \cite{Rocha-Moran:2018jzu}.


{
\renewcommand{\arraystretch}{1.4}
\begin{table}[t]
\centering
\begin{tabular}{ccc} 
\hline \hline 
Field & Group & Coupling \\ 
 \hline 
$B$ & $U(1)_Y$ & $g_1$ \\ 
$W$ & $SU(2)_L$ & $g_2$\\ 
$g$ & $SU(3)_c$ & $g_3$\\ 
$B_X$ & $U(1)_X$ & $g_X$ \\
\hline \hline
\end{tabular} 
\caption{Gauge sector of the model of \cite{Sierra:2015fma}.}
\label{tab:DarkBS1}
\end{table}

\begin{table}
\centering
\begin{tabular}{ccccc} 
\hline \hline 
Field & Spin  & \( SU(3)_c \times\, SU(2)_L \times\, U(1)_Y \times U(1)_X \) \\ 
\hline
\(\phi\) & \(0\)  & \(({\bf 1}, {\bf 1}, 0, 2) \) \\ 
\(\chi\) & \(0\)  & \(({\bf 1}, {\bf 1}, 0, -1) \) \\ 
\(Q_{L,R}\) & \(\frac{1}{2}\)  & \(({\bf 3}, {\bf 2}, \frac{1}{6}, 2) \) \\ 
\(L_{L,R}\) & \(\frac{1}{2}\)  & \(({\bf 1}, {\bf 2}, -\frac{1}{2}, 2) \) \\ 
\hline \hline
\end{tabular} 
\caption{New scalars and fermions in the model of \cite{Sierra:2015fma}.}
\label{tab:DarkBS2}
\end{table}
}


The model extends the SM gauge group with a new dark $U(1)_X$ factor,
under which all the SM particles are assumed to be singlets. The {\it
  dark sector} contains two pairs of vector-like fermions, $Q$ and
$L$, as well as the complex scalar fields, $\phi$ and $\chi$. Tables
\ref{tab:DarkBS1} and \ref{tab:DarkBS2} show all the details about the
gauge sector and the new scalars and fermions in the model. $Q$ and
$L$ have the same representation under the SM gauge group as the SM
doublets $q$ and $\ell$, and they can be decomposed under $SU(2)_L$ as
\begin{equation} \label{eq:decomp}
Q_{L,R} = \left( \begin{array}{c}
U \\
D \end{array} \right)_{L,R} \quad , \quad L_{L,R} = \left( \begin{array}{c}
N \\
E \end{array} \right)_{L,R} \, ,
\end{equation}
with the electric charges of $U$, $D$, $N$ and $E$ being $+2/3$,
$-1/3$, $0$ and $-1$, respectively. In contrast to their SM
counterparts, $Q$ and $L$ are vector-like fermions charged under the
dark $U(1)_X$. In addition to the usual canonical kinetic terms, the
new vector-like fermions $Q$ and $L$ have Dirac mass terms,
\begin{equation} \label{eq:VectorMass}
\mathcal L_m = m_Q \, \overline Q Q + m_L \, \overline L L \, ,
\end{equation}
as well as Yukawa couplings with the SM doublets $q$ and $\ell$ and
the scalar $\phi$,
\begin{equation} \label{eq:VectorYukawa}
\mathcal L_Y = \lambda_Q \, \overline{Q_R} \, \phi \, q_L + \lambda_L \, \overline{L_R} \, \phi \, \ell_L + \hc \, ,
\end{equation}
where $\lambda_Q$ and $\lambda_L$ are $3$ component vectors. The
scalar potential of the model can be split into different pieces,
\begin{equation}
\mathcal V = \mathcal V_{\text{SM}} + \mathcal V\left( H , \phi , \chi \right) + \mathcal V\left( \phi , \chi \right) \, .
\end{equation}
$\mathcal V_{\text{SM}}$ is the usual SM scalar potential containing
quadratic and quartic terms for the Higgs doublet $H$. The new terms
involving the scalars $\phi$ and $\chi$ are
\begin{equation}
\mathcal V\left( H , \phi , \chi \right) = \lambda_{H \phi} \, |H|^2 |\phi|^2 + \lambda_{H \chi} \, |H|^2 |\chi|^2
\end{equation}
and
\begin{equation}
\mathcal V\left( \phi , \chi \right) = m_\phi^2 |\phi|^2 + m_\chi^2 |\chi|^2 + \frac{\lambda_\phi}{2} |\phi|^4 + \frac{\lambda_\chi}{2} |\chi|^4 + \lambda_{\phi \chi} \, |\phi|^2 |\chi|^2 + \left( \mu \, \phi \chi^2 + \hc \right) \, .
\end{equation}
All $\lambda_i$ couplings are dimensionless, whereas $\mu$ has
dimensions of mass and $m_\phi^2$ and $m_\chi^2$ have dimensions of
mass$^2$. We will assume that the scalar potential parameters allow
for the vacuum configuration
\begin{equation}
\langle H^0 \rangle = \frac{v}{\sqrt{2}} \qquad , \qquad \langle \phi \rangle = \frac{v_\phi}{\sqrt{2}} \, ,
\end{equation}
where $H^0$ is the neutral component of the Higgs doublet $H$. The
scalar $\chi$ does not get a vacuum expectation value
(VEV). Therefore, the scalar $\phi$ will be responsible for the
spontaneous breaking of $U(1)_X$. This automatically leads to the
existence of a new massive gauge boson, the $Z^\prime$ boson, with
mass $m_{Z^\prime} = 2 g_X v_\phi$. In the absence of mixing between
the $U(1)$ gauge bosons, the $Z^\prime$ boson can be identified with
the original $B_X$ boson in Table \ref{tab:DarkBS1}. We note that a
Lagrangian term of the form $\mathcal L \supset \varepsilon \, F_{\mu
  \nu}^Y F^{\mu \nu}_X$, where $F_{\mu \nu}^{X,Y}$ are the usual field
strength tensors for the $U(1)_{X,Y}$ groups, would induce this
mixing. In order to avoid phenomenological difficulties associated to
this mixing we will assume that $\varepsilon \ll 1$. Moreover, it can
be shown that loop contributions to this mixing are kept under control
if $m_Q \simeq m_L$ \cite{Sierra:2015fma}.

\begin{figure}
\centering
\includegraphics[scale=0.5]{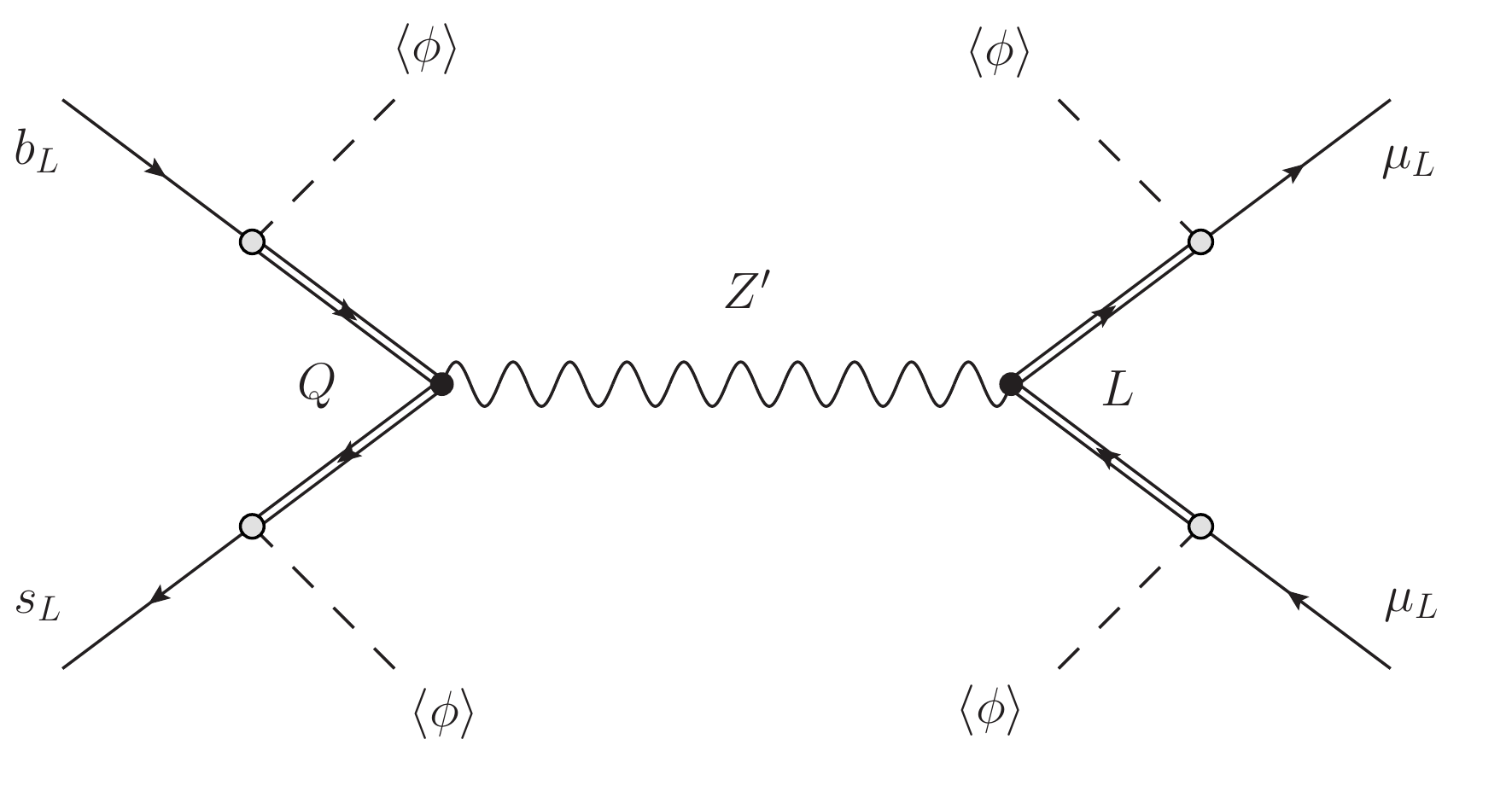}
\caption{Generation of $\mathcal O_9$ and $\mathcal O_{10}$ in the
  model of \cite{Sierra:2015fma}. The mixing between the SM fermions
  and the vector-like ones induce semileptonic four-fermion
  interactions.}
\label{fig:couplings}
\end{figure}

Let us now discuss how this model solves the {\bf $\boldsymbol{b \to
    s}$ anomalies}. After the spontaneous breaking of $U(1)_X$, the
Yukawa interactions in Eq. \eqref{eq:VectorYukawa} lead to mixings
between the vector-like fermions and their SM counterparts. This
mixing results in $Z^\prime$ effective couplings to the SM
fermions. If these are parametrized as
\cite{Buras:2012jb,Altmannshofer:2014rta}
\begin{equation}
\mathcal L \supset \bar f_i \gamma^\mu \left( \Delta_L^{f_i f_j} P_L + \Delta_R^{f_i f_j} P_R \right) f_j Z_\mu^\prime \, ,
\end{equation}
and one assumes $\lambda_Q^d = \lambda_L^e = \lambda_L^\tau = 0$ for
the sake of simplicity, the $Z^\prime$ couplings to $\overline b s$
and $\mu^+ \mu^-$, necessary to solve the $b \to s$ anomalies, are
found to be
\begin{equation}
\Delta_L^{bs} = \frac{2 \, g_X  \lambda_Q^{b} \lambda_Q^{s \ast} v_\phi^2}{2 m_Q^2 + \left(|\lambda_Q^{s}|^2+|\lambda_Q^{b}|^2\right) v_\phi^2} \quad , \quad
\Delta_L^{\mu \mu} = \frac{2 \, g_X |\lambda_L^{\mu}|^2 v_\phi^2}{2 m_L^2 + |\lambda_L^{\mu}|^2 v_\phi^2} \, .
\end{equation}
These couplings induce a tree-level contribution to the semileptonic
four-fermion operators in Eqs. \eqref{eq:O9} and \eqref{eq:O10}, as
shown esquematically in Fig. \ref{fig:couplings}. More specifically,
given that the SM fermions participating in the effective vertices are
left-handed, see Eq. \eqref{eq:VectorYukawa}, the operators $\mathcal
O_9$ and $\mathcal O_{10}$ are generated simultaneously,
with~\cite{Altmannshofer:2014rta}
\begin{equation} \label{eq:c9c10}
C_9^{\mu , \text{NP}} = - C_{10}^{\mu , \text{NP}} = - \frac{\Delta_L^{bs} \Delta_L^{\mu \mu}}{V_{tb} V_{ts}^\ast} \, \left( \frac{\Lambda_v}{m_{Z^\prime}} \right)^2 \, ,
\end{equation}
where we have introduced
\begin{equation} \label{eq:Lambdav}
\Lambda_v = \left( \frac{\pi}{\sqrt{2} G_F \alpha} \right)^{1/2}
\simeq 4.94 \, \text{TeV} \, ,
\end{equation}
with $\alpha = \frac{e^2}{4 \pi}$ the electromagnetic fine structure
constant. $\Lambda_v$ and the CKM elements appear in
Eq. \eqref{eq:c9c10} in order to normalize the Wilson coefficients as
defined in Eqs. \eqref{eq:O9} and \eqref{eq:O10}. By taking proper
ranges for the model parameters, the required values for these Wilson
coefficients, previously identified by the global fits to flavor data,
can be easily obtained.

\begin{figure}
 \centering
 \includegraphics[width=0.6\linewidth]{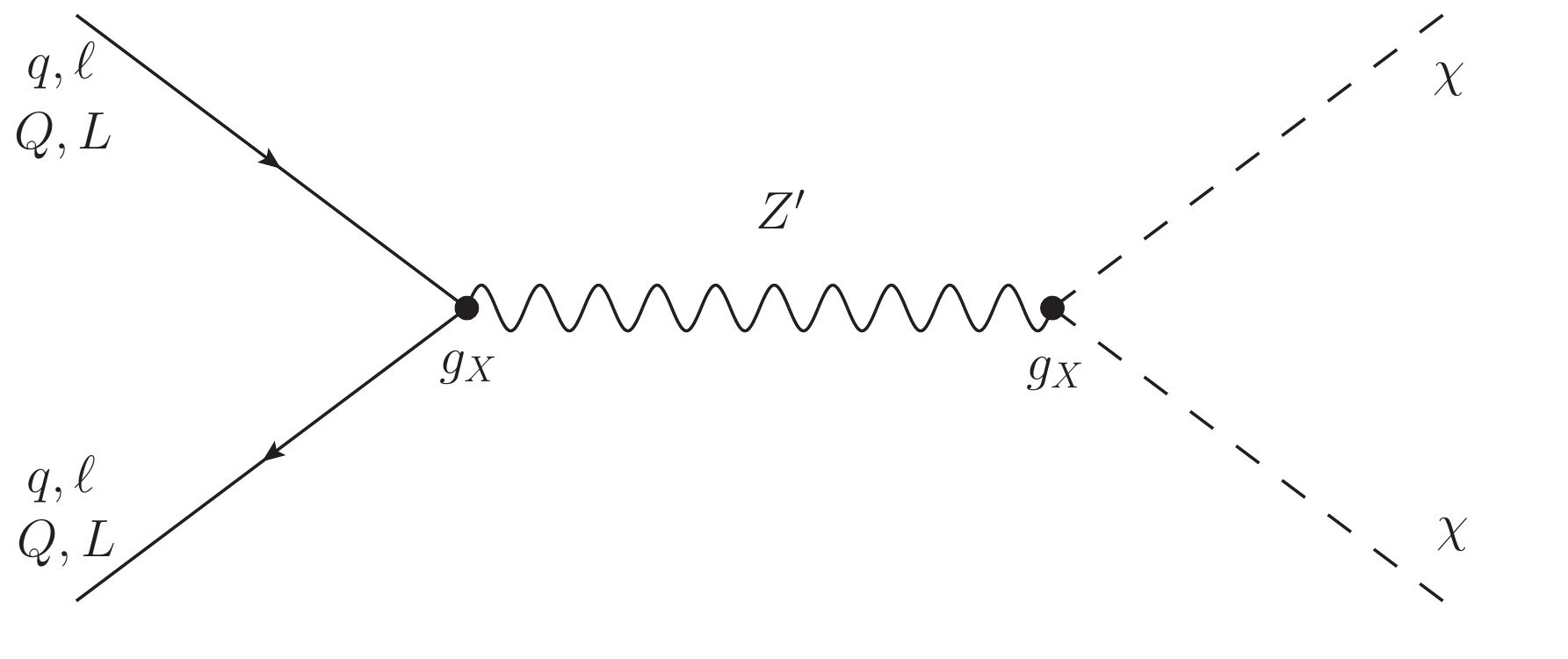}
 \caption{DM production via the $Z^\prime$ portal in the model of \cite{Sierra:2015fma}. We notice that the vertex on the left of the diagram also participates in the explanation of the $b \to s$ anomalies (see Fig.\ref{fig:couplings}).}
 \label{fig:ZprimePortal}
 \end{figure}

\begin{figure}
 \centering
 \includegraphics[width=0.5\linewidth]{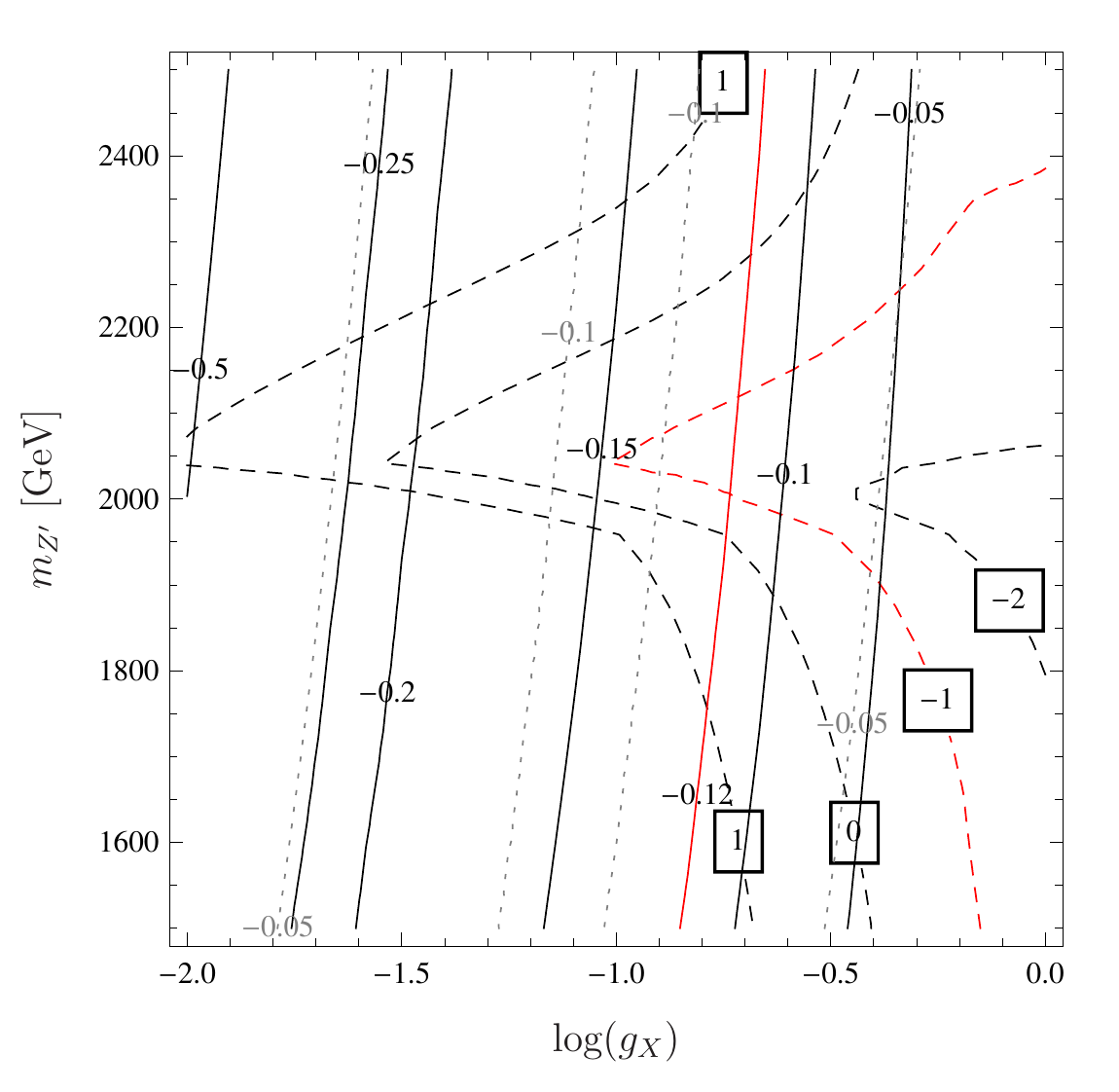}
 \caption{Contours for constant
   $C_9^{\mu,\text{NP}}/C_9^{\mu,\text{SM}}$ and
   $\log(\Omega_{\text{DM}} h^2)$ (dashed black) in the $(g_X,
   m_{Z^\prime})$ plane. For the ratio
   $C_9^{\mu,\text{NP}}/C_9^{\mu,\text{SM}}$ the full 1-loop results
   are shown via the black lines, while the dotted grey lines give the
   values using the tree-level approximation. Red lines indicate the
   preferred values for $C_9^{\mu,\text{NP}}/C_9^{\mu,\text{SM}}$ and
   $\log(\Omega_{\text{DM}} h^2)$ from global fits and cosmological
   observations, respectively. Figure taken from
   \cite{Sierra:2015fma}.}
 \label{fig:C9DM}
 \end{figure}

Finally, we move to the discussion on the {\bf Dark Matter}
phenomenology of the model. We note that the model does not include
any {\it ad-hoc} stabilizing symmetry for the DM candidate
$\chi$. However, this state is perfectly stable. This is due to the
fact that the continuous $U(1)_X$ symmetry leaves a remnant
$\mathbb{Z}_2$ parity, under which $\chi$ is odd, after spontaneous
symmetry breaking
\cite{Krauss:1988zc,Petersen:2009ip,Sierra:2014kua}. Therefore, the
same symmetry that leads to the dynamics behind the $b \to s$
anomalies is also at the origin of the DM stabilization
mechanism. Furthermore, the DM production in the early Universe can
take place via $2\leftrightarrow 2$ processes mediated by the
$Z^\prime$ boson, thus establishing another link with the $b \to s$
anomalies. Indeed, purely gauge interactions open a {\it $Z^\prime$
  portal} that induces $\bar F F\leftrightarrow \chi\chi^*$
annihilation processes, with $F = q, \ell, Q, L$, as shown in
Fig. \ref{fig:ZprimePortal}.~\footnote{Another possibility is the
  so-called {\it Higgs portal}, activated in this model with the
  scalar potential term $\lambda_{H \chi} \, |H|^2 |\chi|^2$, which
  induces $HH^\dagger \leftrightarrow \chi\chi^*$ processes. This DM
  production mechanism will be subdominant for sufficiently small
  $\lambda_{H \chi}$.} We notice that these processes match those in
Fig. \ref{fig:couplings} if one trades one of the fermion pairs for
$\chi\chi^*$. Therefore, one can establish an interplay between flavor
and DM physics in this scenario. Fig.~\ref{fig:C9DM} illustrates this
connection displaying contours for constant $\log(\Omega_{\text{DM}}
h^2)$ (the DM relic density) and the ratio
$C_9^{\mu,\text{NP}}/C_9^{\mu,\text{SM}}$ in the $(g_X, m_{Z^\prime})$
plane. This figure has been obtained with fixed $\lambda_Q^b =
\lambda_Q^s=0.025$, $\lambda_L^\mu = 0.5$, $m_Q=m_L=1~\text{TeV}$ and
$m^2_\chi = 1~\text{TeV}^2$. The calculation of the flavor observables
has been performed with {\tt FlavorKit} \cite{Porod:2014xia}, whereas
the DM relic density has been evaluated with {\tt MicrOmegas}
\cite{Belanger:2013oya}. We see that there is a region in parameter
space, with moderately large $g_X \simeq 0.3$, where the observed DM
relic density can be reproduced and a ratio
$C_9^{\mu,\text{NP}}/C_9^{\mu,\text{SM}}$ in agreement with the global
fits is obtained. We also note that the DM relic density tends to be
large. In fact, in order to obtain a numerical value in the ballpark
of $\Omega_{\rm DM} h^2 \simeq 0.1$ one has to be rather close to the
resonant region with $m_{Z^\prime} \simeq 2 \, m_\chi$, which in this
plot is located around $m_{Z^\prime} = 2$ TeV.

Besides flavor and DM physics, the model has rich phenomenological
prospects in other fronts. The new states can be discovered at the LHC
in large portions of the parameter space. Although one typically
assumes that the $Z^\prime$ boson couples predominantly to the second
and third generation quarks ($|\lambda_Q^d| \ll 1$), the resulting
suppressed production cross-sections at the LHC can still be
sufficient for a discovery, see for instance
\cite{Greljo:2015mma}. Furthermore, the new VL fermions can also be
produced and detected. In particular, the heavy VL quarks masses are
already pushed beyond the TeV scale due to their efficient production
in $pp$ collisions. In what concerns direct and indirect DM detection,
scenarios with a \textit{dark} $Z^\prime$ portal have been discussed
in \cite{Alves:2013tqa,Alves:2015pea}. For more details about this
model, its predictions and the most relevant experimental constraints
we refer to \cite{Sierra:2015fma}.

\section{An example loop model} \label{sec:model2}

We now turn our attention to the second class of models, those that
explain the $b \to s$ anomalies via loop diagrams including DM
particles. A simple but illustrative example of this category is that
presented in \cite{Kawamura:2017ecz}. Previous work on loop models for
the $b \to s$ anomalies, without connecting to the DM problem, can be
found in \cite{Gripaios:2015gra,Arnan:2016cpy}.


{
\renewcommand{\arraystretch}{1.4}
\begin{table}[t]
\centering
\begin{tabular}{cccccc} 
\hline \hline 
Field & Spin  & \( SU(3)_c \times\, SU(2)_L \times\, U(1)_Y \) & \( U(1)_X \) \\ 
\hline
\(X\) & \(0\)  & \(({\bf 1}, {\bf 1}, 0) \) & \(-1\) \\ 
\(Q_{L,R}\) & \(\frac{1}{2}\)  & \(({\bf 3}, {\bf 2}, \frac{1}{6}) \) & \(\phantom{+}1\) \\ 
\(L_{L,R}\) & \(\frac{1}{2}\)  & \(({\bf 1}, {\bf 2}, -\frac{1}{2}) \) & \(\phantom{+}1\) \\ 
\hline \hline
\end{tabular} 
\caption{New scalars and fermions in the model of \cite{Kawamura:2017ecz}. The $U(1)_X$ symmetry is global.}
\label{tab:loopModel}
\end{table}
}

The model introduces two VL fermions, $Q$ and $L$, with the same gauge
quantum numbers as the SM quark and lepton doublets, respectively. It
also adds the complex scalar $X$, a complete singlet under the SM
gauge symmetry. The new fields are charged under a global Abelian
symmetry, $U(1)_X$, under which all SM fields are assumed to be
singlets. As we will see below, this particle content is sufficient to
address the $b \to s$ anomalies. Table \ref{tab:loopModel} details the
new fields and their charges under the gauge and global symmetries of
the model.

The VL fermions $Q$ and $L$ can be decomposed as in
Eq. \eqref{eq:decomp}, with their $SU(2)_L$ components having exactly
the same electric charges as in that case. Therefore, the same Dirac
mass terms as in Eq. \eqref{eq:VectorMass} can be written. In
addition, the symmetries of the model allow for the Yukawa couplings
with the SM doublets $q$ and $\ell$ and the scalar $X$
\begin{equation} \label{eq:VectorYukawa2}
\mathcal L_Y = \lambda_Q \, \overline{Q_R} \, X \, q_L + \lambda_L \, \overline{L_R} \, X \, \ell_L + \hc \, ,
\end{equation}
where $\lambda_Q$ and $\lambda_L$ are $3$ component vectors. The
scalar potential of the model contains the following terms
\begin{equation}
\mathcal V = \mathcal V_{\text{SM}} + m_X^2 |X|^2 + \lambda_X \, |X|^4 + \lambda_H \, |H|^2 |X|^2 \, .
\end{equation}
All $\lambda_i$ couplings are dimensionless, whereas $m_X^2$ has
dimensions of mass$^2$. In the following, possible effects due to the
$\lambda_H$ coupling will be ignored, assuming $\lambda_H \ll 1$. We
will also assume that the scalar potential parameters allow for a
vacuum configuration with $\langle X \rangle = 0$. In this case, the
global $U(1)_X$ symmetry is conserved and the lightest state with a
non-vanishing charge under this symmetry is completely
stable. Moreover, we note that the conservation of $U(1)_X$ prevents
the VL fermions from mixing with the SM ones.

\begin{figure}
\centering
\includegraphics[scale=0.7]{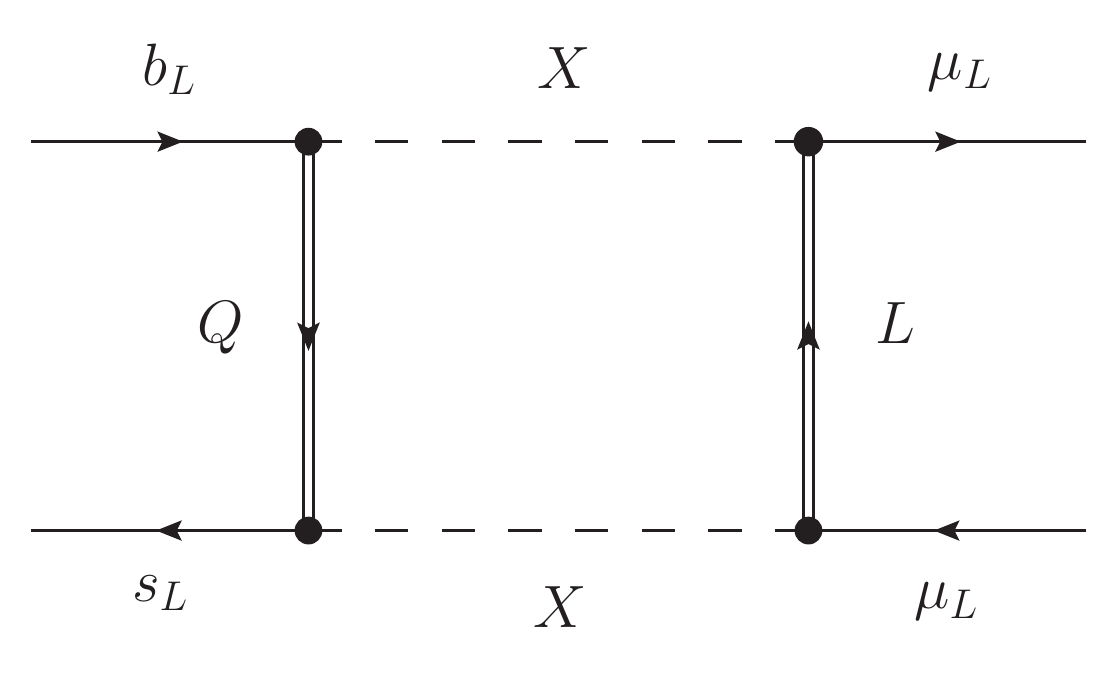}
\caption{Generation of $\mathcal O_9$ and $\mathcal O_{10}$ in the
  model of \cite{Kawamura:2017ecz}. Semileptonic four-fermion
  operators are generated at the 1-loop level.}
\label{fig:loop}
\end{figure}

\begin{figure}
 \centering
 \includegraphics[width=0.6\linewidth]{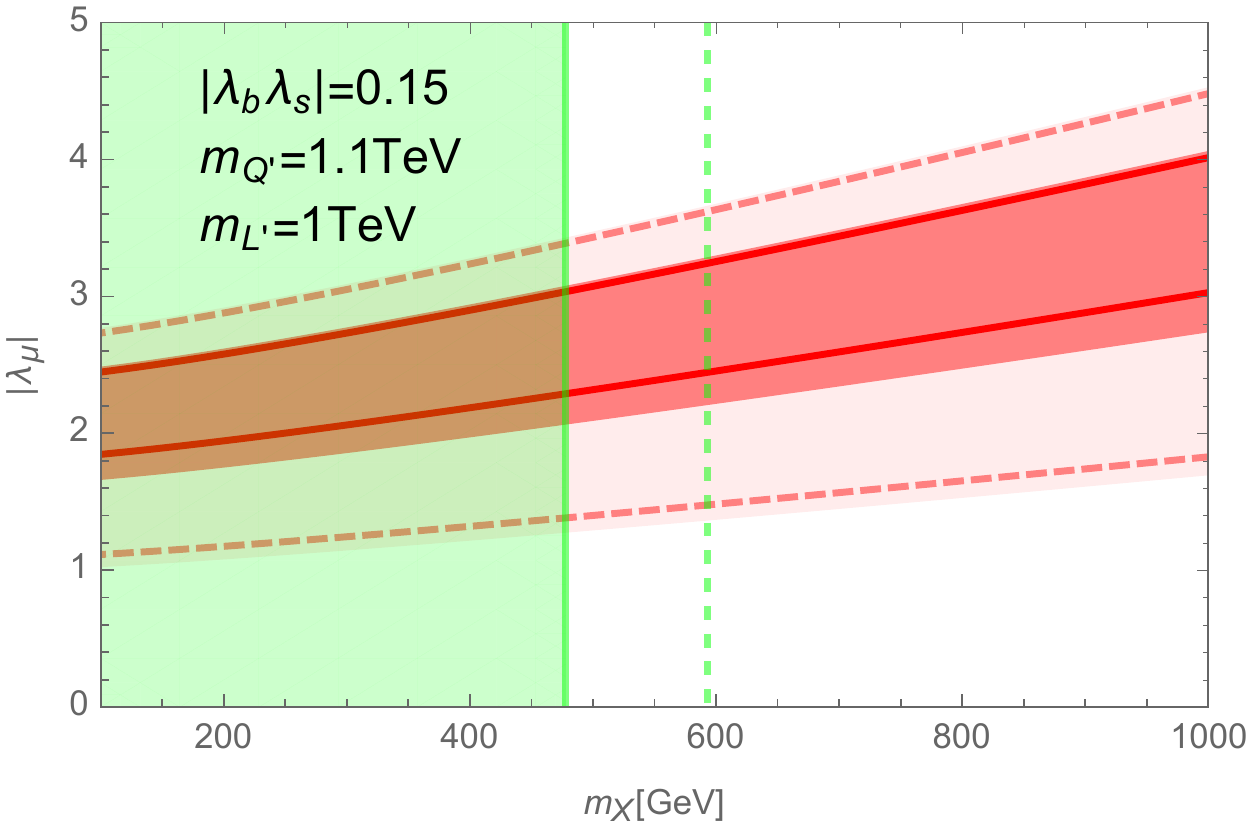}
 \caption{Required values for $\lambda_L^\mu$ (denoted as
   $\lambda_\mu$ in this figure) and $m_X$ to explain the observed
   values of $R_K$ and $R_{K^\ast}$ in the model of
   \cite{Kawamura:2017ecz}. This figure has been obtained with fixed
   $|\lambda_Q^b \lambda_Q^s| = 0.15$, $m_L = 1$ TeV and $m_Q = 1.1$
   TeV. The light (dark) red region corresponds to the $R_K$
   measurement at $1 \sigma$ ($2 \sigma$), whereas the red lines
   indicate the same regions for $R_{K^\ast}$. The green region is
   excluded due to $B_s - \overline{B_s}$ mixing for $m_Q = 1.1$
   TeV. The excluded region would extend up to the dashed green line
   for $m_Q = 1$ TeV. Figure taken from \cite{Kawamura:2017ecz}.}
 \label{fig:LoopPhenoBS}
\end{figure}

We move now to the solution of the {\bf $\boldsymbol{b \to s}$
  anomalies} in the context of this model. It is straightforward to
check that no NP contributions to $b \to s$ transitions are generated
at tree-level in this model.~\footnote{For instance, in constrast to
  the model discussed in Sec. \ref{sec:model1}, there is no SM-VL
  mixing, nor a $Z^\prime$ boson that can mediate these transitions at
  tree-level.} However, the semileptonic operators $\mathcal O_9$ and
$\mathcal O_{10}$ are generated at the 1-loop level as shown in
Fig. \ref{fig:loop}. This diagram leads to
\begin{equation} \label{eq:c9c102}
C_9^{\mu , \text{NP}} = - C_{10}^{\mu , \text{NP}} = \frac{\lambda_Q^b \lambda_Q^{s \ast} |\lambda_L^{\mu}|^2}{64 \, \pi^2 \, V_{tb} V_{ts}^\ast} \, \frac{\Lambda_v^2}{m_Q^2 - m_L^2} \, \left[ f\left( \frac{m_X^2}{m_Q^2} \right) - f\left( \frac{m_X^2}{m_L^2} \right) \right] \, ,
\end{equation}
where $\Lambda_v$ was introduced in Eq. \eqref{eq:Lambdav} and $f(x)$
is the loop function
\begin{equation} \label{eq:loopfunction}
f(x) = \frac{1}{x-1} - \frac{\ln x}{(x-1)^2} \, .
\end{equation}
This loop-level solution to the $b \to s$ anomalies corresponds to
scenario A-I, model class b), in
\cite{Arnan:2016cpy}. Fig. \ref{fig:LoopPhenoBS} shows that the model
can accommodate the $R_K$ and $R_{K^\ast}$ measurements by the LHCb
collaboration. This figure has been obtained with fixed $|\lambda_Q^b
\lambda_Q^s| = 0.15$, $m_L = 1$ TeV and $m_Q = 1.1$ TeV. One finds
that the $1$ and $2 \sigma$ regions for $R_K$ and $R_{K^\ast}$ almost
overlap, and thus they can be accommodated in the same region of
parameter space. Furthermore, in order to be compatible with the
bounds coming from $B_s - \overline{B_s}$ mixing one needs
$|\lambda_Q^b \lambda_Q^s| \ll 1$, which implies a relatively large
value of $|\lambda_L^\mu|$, $|\lambda_L^\mu| \gtrsim 2$. This feature,
a hierarchy between the NP couplings to quarks and leptons, is shared
by most models addressing the $b \to s$ anomalies. For a general
discussion about the $B_s - \overline{B_s}$ mixing constraint in the
context of the $b \to s$ anomalies we refer to \cite{DiLuzio:2017fdq}.

\begin{figure}
 \centering
 \includegraphics[width=0.5\linewidth]{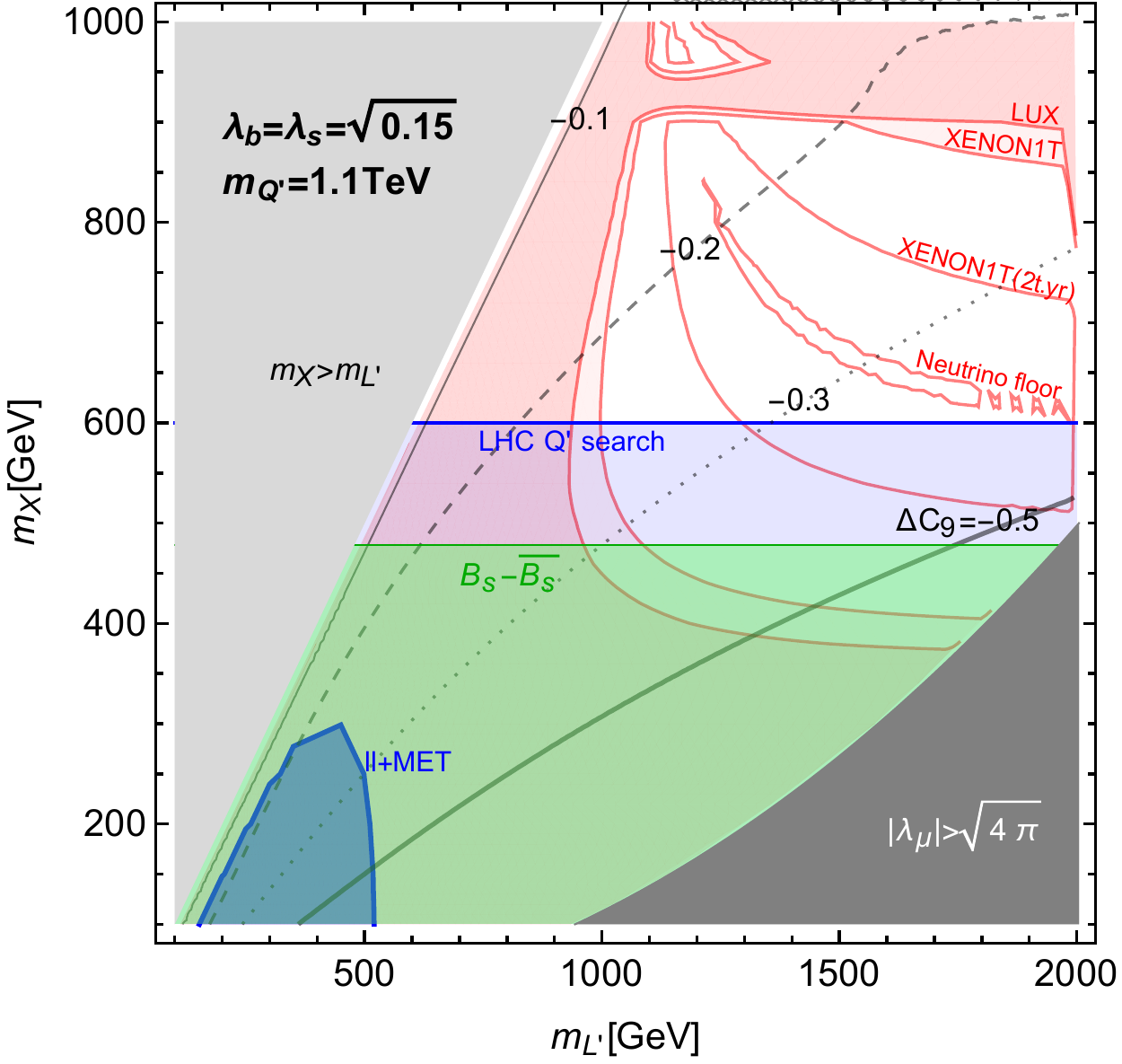}
 \caption{Contours of constant $C_9^{\mu , \text{NP}}$ in the $m_L -
   m_X$ plane for the model of \cite{Kawamura:2017ecz}. This figure
   has been obtained with fixed $\lambda_Q^b = \lambda_Q^s =
   \sqrt{0.15}$ and $m_Q = 1.1$ TeV, choosing $\lambda_L^\mu$ in order
   to reproduce the observed DM relic density. The colored regions are
   excluded by various constraints: heavy quark and lepton searches at
   the LHC (blue), $B_s - \overline{B_s}$ mixing (green) and direct DM
   detection experiments (red). The grey regions are excluded due to
   perturbativity constraints (dark grey region) or by demanding that
   $X$ is the lightest $U(1)_X$-charged state (light grey
   region). Future direct DM detection prospects are also shown in
   this plot. Figure taken from \cite{Kawamura:2017ecz}.}
 \label{fig:LoopPhenoDM}
\end{figure}

Finally, let us discuss the {\bf Dark Matter} phenomenology of the
model. As explained above, the global $U(1)_X$ symmetry is assumed to
be conserved, and this implies that a stable state must
exist. Assuming that the lighest state charged under $U(1)_X$ is the
neutral scalar $X$, it constitutes the DM candidate in the model. One
then needs to determine whether the observed DM relic density can be
achieved in the region of parameter space where the $b \to s$
anomalies are solved, without conflict with other experimental
constraints. This is shown in Fig. \ref{fig:LoopPhenoDM}, where
contours of $C_9^{\mu , \text{NP}}$ are shown in the $m_L - m_X$
plane.  This figure has been obtained with fixed $\lambda_Q^b =
\lambda_Q^s = \sqrt{0.15}$ and $m_Q = 1.1$ TeV. For each parameter
point, the value of $\lambda_L^\mu$ is chosen to reproduce the
observed DM relic density, which is calculated using {\tt MicrOmegas}
\cite{Belanger:2013oya}. Large values of $|\lambda_L^\mu|$ are
obtained in this way. For this reason, the most relevant DM
annihilation channels for the determination of the relic density with
these parameter values are $X X^\ast \leftrightarrow \mu^+ \mu^-, \nu
\nu$. Even though the experimental constraints, in particular those
from direct LHC searches for extra quarks, reduce the allowed
parameter space substantially, one finds valid regions with $C_9^{\mu
  , \text{NP}} \sim -0.3$. This value would explain the $b \to s$
anomalies at $2 \sigma$, see for instance
\cite{Altmannshofer:2017yso}. Interestingly, the model is testable in
future direct DM detection experiments, such as XENON1T, as shown in
Fig. \ref{fig:LoopPhenoDM}. In the region of parameter space selected
for this Figure, the dominant process leading to DM-nucleon scattering
is 1-loop photon exchange, with leptons running in the loop. The loop
suppression is compensated by the large $\lambda_L^\mu$ coupling.

The new states in this model can be discovered at the LHC. For
instance, the heavy VL charged lepton can be produced in Drell-Yan
processes. Due to the required large values for the $\lambda_L^\mu$
coupling, this exotic state is expected to decay mainly to a DM
particle $X$ (invisible at the LHC) and a muon. Since $U(1)_X$
conservation requires the $X$ particles to be produced in pairs, the
expected signature is the observation of two energetic muons and
large missing energy. Similar events replacing the muons by jets
(mainly $b$ jets) are expected for the VL quarks. We conclude the
discussion of this model by referring for more details to the original
work in \cite{Kawamura:2017ecz}.

\section{Summary and discussion} \label{sec:conclusions}

In this mini-review we have discussed New Physics models that address
the $b \to s$ anomalies and link them to the dark matter of the
Universe. The interplay between these two areas of particle physics
may offer novel model building directions as well as additional
phenomenological tests for the proposed scenarios. We have shown that
most of the proposed models can be classified into two categories: (i)
models in which the $b \to s$ anomalies and the DM production
mechanism share a common mediator (such as a heavy $Z^\prime$ boson),
and (ii) models that induce the NP contributions to explain the $b \to
s$ anomalies via loops including the DM particle. These generic ideas
have been illustrated with two particular realizations (the models
introduced in \cite{Sierra:2015fma} and \cite{Kawamura:2017ecz}),
which clearly show that the combination of flavor physics and dark
matter leads to new scenarios with a rich phenomenology.

The introduction of a dark sector in a model for the $b \to s$
anomalies can also have phenomenological consequences besides the
existence of a DM candidate. For instance, both problems, the dark
matter of the Universe and the $b \to s$ anomalies, might be connected
to another long-standing question in particle physics: the muon
anomalous magnetic moment
\cite{Allanach:2015gkd,Belanger:2016ywb,DiChiara:2017cjq,Kowalska:2017iqv}. Furthermore,
it is interesting to note that the introduction of dark matter in some
models can help alleaviating some of the most stringent
constraints. Indeed, the LHC bounds on some mediators become weaker if
they have invisible decay channels~\cite{Faroughy:2016osc}. We believe
that this is a promising line of research to be pursued in order to
fully assess the validity of some scenarios that are currently under
experimental tension.

We are living an exciting moment in flavor physics, with several
interesting anomalies in B-meson decays. Whether real or not, only
time can tell. New LHCb analyses based on larger datasets are expected
to appear in the near future, possibly shedding new light on these
anomalies. In the longer term, fundamental contributions from the
Belle II experiment will also be crucial to settle the issue
\cite{Albrecht:2017odf}. In the meantime, an intense model building
effort is opening new avenues with rich phenomenological
scenarios. The possible connection to one of the central problems in
current physics, the nature of the dark matter of the Universe, would
definitely be a fascinating outcome of this endeavour.

\section*{Acknowledgements}

I am grateful to Junichiro Kawamura, Shohei Okawa and Yuji Omura for
clarifications regarding \cite{Kawamura:2017ecz} and to Javier Virto
for comments about the manuscript. I am also very grateful to my
collaborators in the subjects discussed in this review and acknowledge
financial support from the grants FPA2017-85216-P and SEV-2014-0398
(MINECO), and PROMETEOII/ 2014/084 (Generalitat Valenciana). The
author declares that there is no conflict of interest regarding the
publication of this paper.

\bibliographystyle{utphys}
\bibliography{BSrefs,extra}

\end{document}